\documentclass[journal]{IEEEtran}

\usepackage{graphicx} 
\usepackage{subfigure} 

\usepackage{booktabs}
\usepackage{xspace}
\usepackage{float}
\usepackage{multirow}
\usepackage{verbatim}
\usepackage{algorithmic, algorithm}
\usepackage{amsbsy,amscd,amsfonts,amsgen,amsmath,amsopn,amssymb,amstext,amsxtra}
\usepackage{amsmath,amssymb}
\usepackage{color}
\usepackage{xspace}
\usepackage{float}
\usepackage{multirow}
\usepackage{verbatim}
\usepackage{algorithmic, algorithm}
\usepackage{amsbsy,amscd,amsfonts,amsgen,amsmath,amsopn,amssymb,amstext,amsxtra}
\usepackage{epstopdf}

\ifCLASSINFOpdf

\else

\fi

\newcommand{{\bhat}}{\hat{\bf{b}}}
\newcommand{{\xhat}}{\hat{\bf{x}}}
\newcommand{{\hhat}}{\hat{\bf{h}}}
\newcommand{{\zhat}}{\hat{\bf{z}}}
\renewcommand{\c}{{\bf c}}

\renewcommand{\b}{{\bf b}}
\newcommand{\s}{{\bf s}}
\newcommand{{\x}}{{\bf x}}

\renewcommand{\v}{{\bf v}}

\renewcommand{\v}{{\bf v}}

\renewcommand{\d}{{\bf d}}

\newcommand{\C}{{\bf C}}
\newcommand{\G}{{\bf G}}

\newcommand{\V}{{\bf V}}
\def \X {{\bf X}}

\newcommand{\B}{{\bf B}}

\newcommand{\D}{{\bf D}}

\newcommand{\A}{{\bf A}}
\newcommand{\W}{{\bf W}}

\newcommand{\Xms}{\X_{\text{-}S}}
\newcommand{\Vms}{\V_{\text{-}S}}

\def \C {{\bf C}}

\def \sub {\mathcal{S}}

\newcommand{\nystrom}{Nystr\"{o}m~}

\newcommand{\ny}{Nystr\"{o}m~}

\newcommand{\Ckp}{{\bf C}_{k+1}}
\newcommand{\ckp}{{\bf c}_{k+1}}
\newcommand{\skp}{{\bf s}_{k+1}}
\newcommand{\qkp}{{\bf q}_{k+1}}
\newcommand{\Wkp}{{\bf W}_{k+1}}
\newcommand{\Xkp}{{\bf X}_{k+1}}
\newcommand{\bkp}{{\bf b}_{k+1}}
\newcommand{\dkp}{{\bf d}_{k+1}}
\newcommand{\kp}{_{k+1}}
\newcommand{\inv}{^{-1}}
\newcommand{\eqb}[1]{\begin{equation}\label{#1}}
\newcommand{\eqe}{\end{equation}}
\DeclareMathOperator*{\argmax}{arg\,max}
\DeclareMathOperator*{\colsum}{colsum}

\newtheorem{THEO}{Theorem}
\newtheorem{LEMM}{Lemma}

\usepackage{color}
\definecolor{linkBlue}{rgb}{.1,.1, .6}
\definecolor{linkGreen}{rgb}{.1,.35, .1}

\def\argmax{\mathop{\arg\,\max}\limits}%

\newcommand{\bitm}{\begin{itemize}}
\newcommand{\eitm}{\end{itemize}}
\newcommand{\beqa}{\begin{eqnarray}}
\newcommand{\eeqa}{\end{eqnarray}}

\newcommand{{\nhood}}{\mathcal{N}}

\newcommand{\R}{\mathbb{R}}
\newcommand{\proj}[2]{{\boldsymbol{\pi}}_{#1}(#2)}

\newcommand{{\coef}}{ {\bf x}}

\renewcommand{\Xi}[1]{{\bf X}^{(#1)}}

\begin{document}

\title{Self-Expressive Decompositions for \\ Matrix Approximation and Clustering}

\author{Eva L. Dyer,~\IEEEmembership{Member,~IEEE,}
        Tom A. Goldstein,~\IEEEmembership{Member, ~IEEE,}
        Raajen Patel,~\IEEEmembership{Student Member,~IEEE,}\\
        Konrad P. K\"ording,~
        and Richard G. Baraniuk,~\IEEEmembership{Fellow,~IEEE}
\thanks{Eva L. Dyer and Konrad P. K\"ording are with the Dept. of Physical Medicine and Rehabilitation and the Dept. of Applied Math at Northwestern University,
Chicago, IL, 60611, USA. Raajen Patel and Richard G. Baraniuk are with the Dept. of Electrical and Computer Eng. at Rice University, Houston, TX, 77005. Tom A. Goldstein is with the Computer Science Dept. at University of Maryland, College Park, MD 20742.}}

\maketitle

\begin{abstract} 
Data-aware methods for dimensionality reduction and matrix decomposition aim to find low-dimensional structure in a collection of data. Classical approaches discover such structure by learning a basis that can efficiently express the collection. Recently, ``self expression'', the idea of using a small subset of data vectors to represent the full collection, has been developed as an alternative to learning. Here, we introduce a scalable method for computing sparse SElf-Expressive Decompositions (SEED). SEED is a greedy method that constructs a basis by sequentially selecting incoherent vectors from the dataset. After forming a basis from a subset of vectors in the dataset, SEED then computes a sparse representation of the dataset with respect to this basis. We develop sufficient conditions under which SEED exactly represents low rank matrices and vectors sampled from a unions of independent subspaces. We show how SEED can be used in applications ranging from matrix approximation and denoising to clustering, and apply it to numerous real-world datasets. Our results demonstrate that SEED is an attractive low-complexity alternative to other sparse matrix factorization approaches such as sparse PCA and self-expressive methods for clustering.
\end{abstract} 


\IEEEpeerreviewmaketitle

\section{Introduction}
Data-driven methods for sparse matrix factorization, such as sparse PCA (SPCA) \cite{zou2006sparse} and dictionary learning \cite{olshausen1996emergence,ksvd}, approximate data vectors as sparse linear combinations of a small set of basis elements. While these simple approaches provide an extremely efficient representation of a dataset, the bases learned often ``mix'' points from different low-dimensional geometric structures and thus lead to degraded classification/clustering performance \cite{mahoney2009cur}.

An alternative approach for revealing low-dimensional structure is to let the data ``express itself''---to represent each element in a dataset in terms of a small subset of samples. {\em Self-expression} has already been successfully used in the context of classification \cite{wright2010sparse}, clustering \cite{elhamifar2009sparse,LRR,DyerJMLR13}, and low-rank matrix approximation \cite{mahoney2009cur,deshpande2006matrix}. 

In contrast to learning a basis, self expression provides a {\em provable} means by which low-dimensional subspace structures can be discovered \cite{elhamifar2009sparse,vidaljournal, soltanolkotabi2012geometric, DyerJMLR13}. The idea underlying self-expressive approaches for clustering, such as sparse subspace clustering (SSC) \cite{vidaljournal} and low-rank representations (LRR) \cite{LRR}, is to represent the dataset in terms of all of the signals in the collection.  For a dataset $\X \in \R^{m \times N}$, containing $N$ data vectors of $m$ dimensions, one computes a representation $\X \approx \X \V$, where $\V \in \R^{N \times N}$ is a sparse matrix with zeros along the diagonal. The resulting sparse matrix $\V$ can be interpreted as an affinity matrix, where data vectors that lie in the same subspace (or cluster) are presumed to use one another in their sparse representations. The dataset is then clustered by applying spectral clustering methods \cite{graphcuts} to the graph Laplacian of $\A = | \V | + |\V^T|$.

While self-expressive methods like SSC and LRR provide a principled segmentation of $\X$ into low-dimensional subspaces (subspace clustering), applying these methods to big datasets is challenging. Both SSC and LRR require the construction and  storage of an $N \times N$ affinity matrix for a dataset of size $N$. Even when low-complexity greedy methods are used to populate the affinity matrix, as in SSC-OMP \cite{DyerJMLR13}, clustering the data requires solving an eigenvalue problem for the entire affinity matrix, which is intractable for large $N.$ As such, the development of efficient solutions for decomposing large datasets is essential for both clustering and discovering low-dimensional structures in the data.

\subsection{Our contributions: SEED}
In this paper, we develop a scalable approach for sparse matrix factorization that is built upon the idea of using samples from the data to ``express itself''. Our approach, which we refer to as a SElf-Expressive Decomposition (SEED), consists of two main steps. In the first step, we select data samples by sequentially selecting columns from $\X$ that are {\em incoherent} (uncorrelated) from columns selected at previous iterations. To do this, we use a method called oASIS (Accelerated Sequential Incoherence Selection) \cite{pateloasis}. oASIS operates on a subset of the Gram matrix $\G = \X^T \X$ and can thus be used to quickly select columns from $\X$ without computing the entire Gram matrix. In the second step, we use the vectors selected in the first step as a basis by which we compute a sparse representation of the dataset using a faster variant of the orthogonal matching pursuit (OMP) method \cite{rubinstein2008efficient}. We describe SEED in detail in Sec.\ \ref{sec:seed} and provide pseudocode in Alg.\ \ref{alg:seed}. 

We demonstrate that SEED provides an effective strategy for matrix approximation, both in theory (Thm.\ \ref{thm:exactrecovery}) and in practice (Fig.\ \ref{fig:mainfig}). In particular, Thm.\ \ref{thm:exactrecovery} provides a sufficient condition for oASIS to return a subset of columns $\X_S$ that captures the full range of the data, i.e., $\X = \X_S \X_S^{+} \X$. This condition, called {\em exact matrix recovery}, highlights two attractive properties of our proposed approach for column selection: (i) it naturally selects linearly independent columns and thus provides a highly efficient representation of low rank matrices, and (ii) it provides an estimate of the representation error remaining in the dataset after each iteration.  This error estimate can be used to stop the algorithm when explicit evaluation of the error is intractable. 

Following our analysis, we demonstrate how SEED can be applied to aid in and/or solve numerous problems including: matrix approximation, clustering, denoising,  and outlier detection (Sec.\ \ref{sec:results}). We evaluate the performance of SEED for these applications on several real-world datasets, including three image datasets and a collection of neural signals from the motor cortex. Our results demonstrate that SEED provides a scalable alternative to other sparse decomposition methods such as SPCA, SSC, and nearest neighbor (NN) methods at a fraction of the computational cost.

\subsection{Paper Organization}
This paper is organized as follows. In Sec.\ \ref{sec:bg} we provide background on column subset selection, sparse recovery, and sparse subspace clustering. In Sec.\ \ref{sec:seed}, we introduce SEED and then provide motivating examples and a complexity analysis. In Sec.\ \ref{sec:exactrecovery}, we develop a sufficient condition for exact matrix recovery with SEED for low rank matrices and datasets living on unions of independent subspaces. In Sec.\ \ref{sec:results}, we study the performance of SEED for four applications: matrix approximation, (ii) denoising, (iii) sparse representation-based learning, and (iv) outlier detection. Finally, we end with concluding remarks (Sec.\ \ref{sec:conclusion}) and further details of our approach for column selection in Appendix A.

\subsection{Notation and Preliminaries}
We denote matrices $\X$ with uppercase bold script and vectors $\x$ with lowercase bold script. We write the $(i,j)$ entry of a matrix $\X$ as $\X_{ij}$. Let $[ \A ~ \B]$ denote the column-wise concatenation of $\A$ and $\B$. Let $\X^{+}$ denote the left pseudoinverse of $\X$. 
The orthogonal projection of $\X$ onto the span of the columns indexed by $S$ is defined as $\proj{S}{\X} = (\X_S \X_S^{+}) \X$. The Frobenius norm is defined as $\| \X \|_F^2 = \sum_{ij} \X_{ij}^2$. The support of a vector $\x$, ${\rm supp}(\x)$, indexes its nonzero elements. The sparsity of $\v$ equals $|{\rm supp}(\x)|$. We denote the columns of $\X$ not indexed by the set $S$ as $\Xms$. We denote entry-wise multiplication of $\A$ and $\B$ by $\A \circ \B$ and ``$\colsum$'' returns the sums of the columns of its argument.

We say that a collection of $N$ signals $\{ \x_i \}_{i = 1}^N$ of dimension $m$ lie on a union of $p$ subspaces $\{  \mathcal{S}_i \}_{i = 1}^p$ in $\R^m$ when each signal $\x_i \in \mathcal{U} = \cup_{i=1}^p \sub_i$, where the dimension of the $i^{\rm th}$ subspace $\sub_i$ equals $k_i < m$. The matrix $\X$ contains independent subspaces when ${\rm rank}(\X) = {\rm dim}(\mathcal{U}) = \sum_{i=1}^p k_i$ and $k_i$ is the subspace dimension. If ${\rm rank}(\X) < \sum_{i=1}^p k_i$, then the subspaces are overlapping.

\section{Background and Related Work}
\label{sec:bg}
\subsection{Column Subset Selection}
Consider a dataset of $N$ vectors in $\R^m,$ each represented by a column of $\X \in \R^{m \times N}.$ The task of identifying $L$ columns that best represent the entire matrix $\X $ is referred to as {\em column subset selection} (CSS). The CSS problem is formulated as follows: 
$$ ({\rm CSS})  \hspace{1cm} \min_{|S| = L} ~~  \|  \X - \proj{S}{\X}  \|_F.$$
The CSS objective aims to find a set of $L$ columns from $\X$ (indexed by the set $S$) that best approximate $\X$ in the least-squares sense. For a collection of signals that lie on a $k$-dimensional subspace, all invertible sub matrices $\X_S \in \R^{m \times k}$ of $\X$ will yield {\em exact matrix recovery}, i.e., 
$$\| \X - \proj{S}{\X} \|_F = 0.$$

Unfortunately,  (CSS) is believed to be NP-hard, since it requires a brute force search for the sub-matrix of $\X$ that provides the best approximation. However, a large body of literature in random and adaptive column selection has emerged over the past few years \cite{journals/jmlr/DrineasM05,deshpande2006matrix}. While uniform random sampling is the easiest and most well-studied sampling method, a number of adaptive selection criteria have been proposed to reduce the number of samples required to achieve a target approximation error. Adaptive approaches include: (i) leverage-based sampling \cite{mahoney2009cur, mahoney2012fast} and (ii) sequential error-based selection (SES) approaches \cite{deshpande2006matrix}. Leverage sampling requires computing a low-rank SVD of the data matrix to determine which columns exert the most influence over its low-rank approximation. After computing the so-called ``leverage scores'' for the columns of the dataset, columns are drawn randomly based upon their leverage score. SES strategies select columns based upon how well they are approximated by the current sample set \cite{deshpande2006matrix}: the probability of selecting the $i^{\rm th}$ column $\x_i$ is proportional to $p(i) \propto \|  \x_i - \proj{S}{\x_i} \|_2$. While SES strategies are highly effective in practice, these methods are very costly, because they require computing a $m \times N$ residual error matrix at each selection step. In contrast, the proposed column selection strategy (oASIS) requires computing and operating on only a $k \times N$ matrix at each step, where $k$ is the iteration number and $k \le L \le m$. 

\subsection{Convex Approach for Subset Selection}
Recently, a convex approach was proposed to find representative columns from $\X$ \cite{elhamifar2012see}. This self-expressive method selects representative columns from $\X$ by solving:
\begin{equation*} 
\min_{\V \in \R^{N \times N}} \quad  \| \V \|_{2,1} \quad {\rm subject ~to} \quad \X = \X\V,
\end{equation*}
where $\| \V \|_{2,1}$ is the sum of the $\ell_2$-norms of the rows of $\V$. By penalizing the rows of $\V$ in this way, we minimize the number of non-zero rows in $\V$ which in turn minimizes the number of columns of $\X$ needed to represent the dataset. This approach is known to reveal representative columns from collections of data \cite{elhamifar2012see,elhamifar2012finding} and also aid in hyperspectral unmixing \cite{fu2013greedy}.  However, this approach requires solving for $N \times N$ matrix and cannot be used directly to enforce sparsity in the entries of $\V$ as in SEED because group sparsity norms are known to produce dense estimates within each group.

\subsection{Greedy Sparse Recovery}
Sparse recovery methods aim to form an approximation $\widehat{\x} = \D \v$ consisting of a small number of nonzero coefficients in $\v$. Greedy methods for sparse recovery, such as orthogonal matching pursuit (OMP) \cite{davisomp}, select columns from $\D$ (atoms) iteratively, subtracting the contribution of each selected atom from the current signal residual. This selection process is then repeated until a stopping criterion is satisfied: either a target sparsity $\|\x\|_0 =k$ is reached (Sparse), or the residual magnitude becomes smaller than a pre-specified value (Error). Pseudocode for the OMP algorithm is given in Alg.\ \ref{alg:omp}. 
 
\begin{algorithm}[t!]
   \caption{: \bf Orthogonal Matching Pursuit (OMP)    \label{alg:omp}}
\begin{algorithmic}
   \STATE {\bfseries Input:} Input signal $\x$, dictionary $\D$ containing $L$ unit-norm vectors in its columns, termination condition (either the target sparsity $k$ or approximation error $\epsilon$).\\
   \vspace{1mm}
   \STATE {\bfseries Output:} A sparse coefficient vector $\v$.
 \vspace{2mm}  
\STATE{\bfseries Initialize:} Set the residual $\bf r$ to the input signal ${\bf r} = \x$.
\STATE{1. Select the column of $\D$ that is maximally correlated with ${\bf r}$ and add it to $\Lambda$ 
$$\Lambda \leftarrow \Lambda ~ \cup ~ \arg \max_{j=1, \dots, N}  | \langle {\bf d}_j{, \bf r} \rangle | .$$}
\STATE{2. Update the residual  ${\bf r} = {\x} - \D_{\Lambda} \D_{\Lambda}^{+} {\x} .$}
\STATE{3. Repeat steps (1--2) until the norm of the residual $\|{\bf r}\| \le \epsilon$ or $|\Lambda| = k$.}
\STATE{4. Return the sparse coefficient vector $\v$, with nonzero coefficients $\v_{\Lambda} = \D_{\Lambda}^{\dagger} \x$ and $\v(i) = 0, ~\forall i \notin \Lambda$.}
\end{algorithmic}
\end{algorithm}

\subsection{Sparse Subspace Clustering}
\label{sec:ssc}
A number of methods for learning multiple subspaces from data (subspace clustering) use the idea of {\em self-expression} to represent the data in terms of other signals in the collection; such methods lead to state-of-the-art clustering performance for unions of subspaces \cite{vidaljournal,LRR}. For instance, sparse subspace clustering (SSC) \cite{vidaljournal} factorizes the dataset $\X \in \R^{M \times N}$ by solving the following $\ell_1$-minimization problem:
$$ \min_{\V \in \R^{N \times N}} ~~ \| \V \|_1 ~~{\rm s.t.}~~ {\rm diag}(\V) = 0,~ \| \X - \X \V \|_F \le \epsilon,$$ where $\epsilon$ is a user set parameter which controls the error in the self-expressive approximation. The idea underlying SSC is that that each datapoint can be represented as a linear combination of a small number of points in the dataset. 

The coefficient matrix $\V$ computed via SSC can be interpreted as a graph, where the $(i,j)$ entry of the matrix represents the edge between the $i^{\rm th}$ and $j^{\rm th}$ point in the dataset; the strength of each edge represents the likelihood that two points live in the same subspace. After forming a symmetric affinity matrix $\A = | \V | + | \V^T |$, spectral clustering is then performed on the graph Laplacian of the affinity matrix to obtain labels (indicating the subspace membership) for all the points in the dataset \cite{graphcuts}. 

The motivation underlying SSC is that the sparse representation of a signal under consideration will consist of other signals from the same subspace. In fact, the SSC procedure leads to provable guarantees of  {\em exact feature selection} (EFS) --- a condition in which every data point is represented using only data from within its own subspace  \cite{vidaljournal, soltanolkotabi2012geometric, DyerJMLR13,wang2013noisy}.  Guarantees for EFS require that there exists at least $k$ linearly independent columns in $\X$ that span each $k$-dimensional subspace in the dataset. When this occurs, we say that $\X$ provides a {\em complete reference set} for a subspace. In Sec.\ \ref{sec:unionsubs}, we show that when $\X$ lies on a union of independent subspaces, SEED can be guaranteed to return a set of columns $\X_S$ that provides a complete reference set for all of the subspaces present in the dataset.

\section{Sparse Self-Expressive Decomposition (SEED)}
\label{sec:seed}
In this Section, we provide a description, pseudocode (Alg.\ \ref{alg:seed}), and complexity analysis of SEED.

\begin{algorithm}[t!]
   \caption{: Sparse Self-Expressive Decomposition (SEED)}
   \label{alg:seed}
\begin{algorithmic}
   \STATE {{\bfseries Input:} A dataset $\X \in \R^{M \times N}$, the maximum number of columns to select $L$, termination criterion for Step 1 $\delta$, a termination criterion for Step 2 (either target sparsity $k$ and/or approximation error $\epsilon$).}\\      
   \STATE {{\bfseries Output:} A normalized basis $\D \in \R^{M \times L}$ and sparse coefficient matrix $\V \in \R^{L \times N}$.}\\
   \vspace{3mm}
\STATE{ {\bf Step I. Column Subset Selection:}~~ Select $L$ columns via oASIS and normalize the selected columns to form $\D \in \R^{M \times L}$.}\\
   \STATE{ {\bf Step II. Greedy Sparse Recovery:}~~ Solve OMP for each column of $\X$ with respect to $\D$ and stack the result into the corresponding column of $\V \in \R^{L \times N}$.}\\
\end{algorithmic}
\end{algorithm}

\subsection{SEED Method}
As we described in the Introduction,  SEED aims to form an approximation $\widehat{\X} \approx \D \V$ such that $\V$ is sparse and $\D$ contains as few normalized columns of $\X$ as necessary to represent the data. Our solution consists of two steps, which we now describe in detail.

\subsubsection{Step 1: Column Selection}
\label{sec:sis}
In Step 1 of SEED, we select columns from $\X$ that form a good low-dimensional approximation to the dataset. To do this, we employ a method called Accelerated Sequential Incoherence Selection (oASIS) \cite{pateloasis}, an adaptive strategy that selects columns that are {\em incoherent} (uncorrelated) from one another. oASIS was originally designed to compute low rank factorizations of positive semidefinite kernel matrices used in a wide range of machine learning applications. Here, we show how oASIS can be used in a novel way, for column selection, by finding a low rank approximation to the Gram matrix $\G = \X^T\X$.  

\begin{figure}[t!]
\centering
\centerline{\includegraphics[width=\columnwidth]{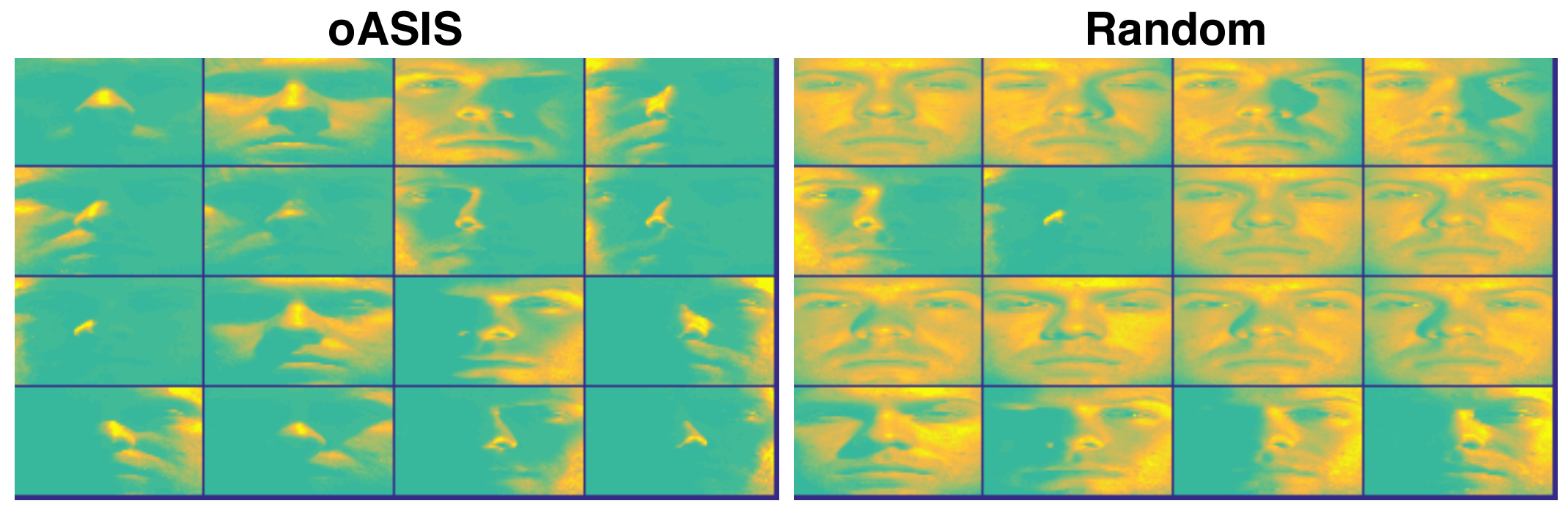}}
\vspace{-0.2cm}
\caption{{\bf Incoherence sampling from face images.} Face images from one subject selected via oASIS (left) and random sampling (right). The images selected via oASIS represent a wide range of diverse illumination conditions whereas random sampling selects a number of similar (redundant) illumination conditions.}\label{fig:facesamples}
\vspace{-3mm}
\end{figure}

As a motivating example, we show the samples (images) selected via oASIS and random sampling for a dataset consisting of faces under various illumination conditions (Fig.\ \ref{fig:facesamples}). We observe that oASIS returns a set of images that are highly varied in terms of their illumination (left) because we select columns that are incoherent from those selected at previous iterations. In contrast, random sampling (right) returns a set of images with highly redundant illumination conditions and is thus less efficient at expressing the range of the dataset. We provide details (Appendix A) and pseudocode (Alg.\ \ref{alg:oasis}) for our implementation of oASIS.

To motivate the selection criterion used in oASIS, suppose that we have already selected $k$ columns from $\X$, indexed by the set $S$. Without loss of generality, we reorder $\X = [\X_S ~ \Xms]$. Let $\W = \X_S^T \X_S$ and $\C = \X^T \X_S$. The least-squares approximation of $\G$ in terms of the subsampled columns $\C$ (i.e., the \ny approximation \cite{Williams01usingthe}) is given by 
$$ \widehat{\G}_k = \C \W^{+} \C^T.$$ 
We assume that $\X_S$ contains linearly independent columns and thus inversion of $\W$ is possible. In Sec.\ \ref{sec:indep}, we will show that oASIS is guaranteed to only select linearly independent columns and thus this assumption is justified.

\begin{figure}[t!]
        \centering
                \includegraphics[width=0.6\columnwidth]{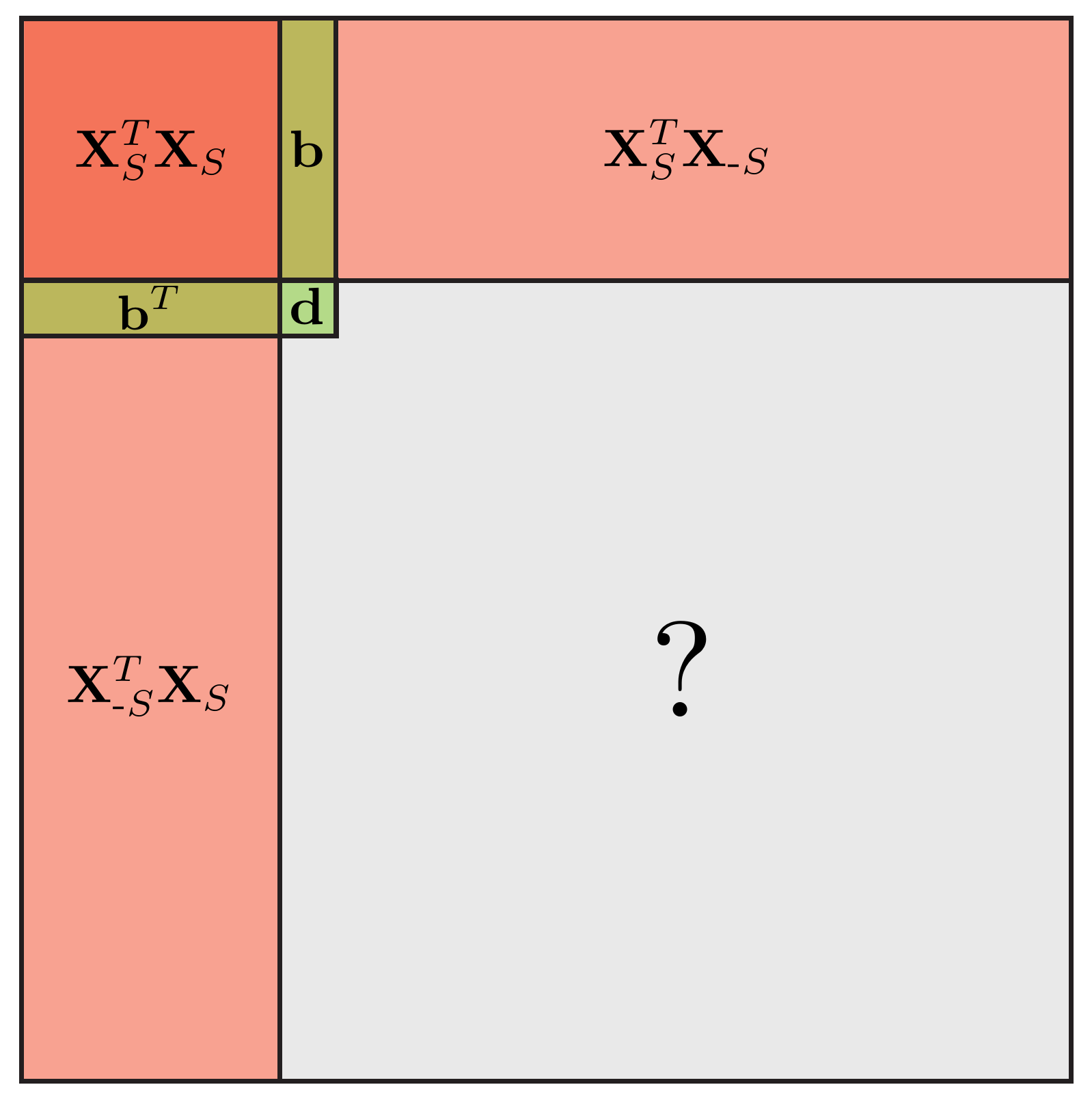}
                \caption{{\bf Column selection with oASIS.} {At each step of oASIS, we project $\b$ (in green) onto the span of $\W$ (upper left red block) and select the column that produces the largest deviation when compared its corresponding diagonal entry $\d$, i.e., we select the column that has maximal deviation $\Delta = | \d - \b^T \W^{-1} \b | $. Our proposed column selection strategy only depends on knowledge of the red shaded regions (a subset of columns/rows of $\G$) and does not depend on the gray shaded region containing the inner products between $\X$ and the unsampled columns $\X_{-S}$ in question.}}
                \label{fig:GMatrix}
 \end{figure}

Now, we would like to determine which column to add to our current approximation $\widehat{\G}_k$ (Fig.\ \ref{fig:GMatrix}). If we consider a new column $\c_i = \X_S^T \x_i$, the resulting upper left $(k+1)$ by $(k+1)$ block of the approximation can be written as 
\begin{align*}
\label{eq:Gkp1}
\widehat{\G}_{k+1} &= \begin{bmatrix}
\Xkp^T \Xkp & \b_i \\
\b_i^T & \d_i
\end{bmatrix}
 \begin{bmatrix} 
\W ^\dagger_k & 0 \\ 
0 & 0 
\end{bmatrix}
  \begin{bmatrix}
\Xkp^T \Xkp & \b_i \\
\b_i^T\ & \d_i
\end{bmatrix}\\
&= \begin{bmatrix}
\X_S^T \X_S & \b_i \\
\b_i^T & \b_i^T \W^{-1} \b_i
\end{bmatrix},
\end{align*}
where $\b_i = \X_S^T \x_i$, $\d_i = \x_i^T \x_i$, and $\X_{k+1} = [\X_S ~ \x_i]$.

If a new column $\b_i$ lies in the span of $\W$, then the projection $\b_i^T \W^{-1} \b = \d_i$. However, the discrepancy between these two scalar quantities provides a measure of how poorly $\W$ represents a candidate column $\c_i$. Thus, without computing the entire column $\c_i$ and measuring its projection onto the span $\W$, we can instead approximate the influence that $\c_i$ will have on the current approximation by measuring the discrepancy between our estimate $\b_i^T \W^{-1} \b$, and $\d_i$. Using this insight, we employ the following greedy strategy to decide which column to add to our approximation at the $k+1$ iteration:
\begin{enumerate}
\item Permute the rows and columns of $\G$ such that the first $k$ columns correspond to the columns that we have already sampled and form  $\W_k^+$. 
\item Let $\b_i$ denote the first $k$ entries of column $i$, and let $\d_i$ denote the diagonal entry in this column. For each unsampled column, calculate the lower bound $\Delta_i$ on how much it changes the \nystrom approximation:
   $$\Delta_i =  \d_i - \b_i^T\W_k^+ \b_i  .$$
\item Select the unsampled column with maximum $| \Delta_i |$ and add it to $S$.  
\item If the selected value of $|\Delta_i|$ is smaller than a user set threshold, then terminate.  Otherwise, let $k \leftarrow k+1,$ and return to Step 1.
 \end{enumerate}

At each iteration, one needs to compute the coefficient matrix (represented by $\W_k^{+}$ above). We can speed up the na\"ive implementation above by performing rank-$1$ updates to the coefficient matrix every time a column is added. For the sake of completeness, we provide a brief description and pseudocode for oASIS in the Appendix. See \cite{pateloasis} for a full discussion of oASIS and its application to low rank kernel matrix approximation. 

\subsubsection{Step 2: Greedy Sparse Recovery}
\label{sec:seedomp}
In Step 2 of SEED, we compute the sparse representations of the columns of $\X$ in terms of the set of columns selected in Step 1, given by $\X_S$. To do this, we first normalize all the columns in $\X_S$ to have unit $\ell_2$-norm; let $\D \in \R^{m \times L}$ denote the corresponding matrix of normalized datapoints. Without loss of generality, we reorder $\X = [ \X_S ~ \Xms]$ and compute its sparse decomposition as $\widehat{\X} = \D \V$, where $ \V = [ {\rm diag}( {\boldsymbol \alpha} ) ~ \W] $ and ${\boldsymbol \alpha} \in \R^{L}$ is a vector containing the $\ell_2$-norm of the $i^{\rm th}$ column in $\X_S$ in its $i^{\rm th}$ entry. The columns of $\W$ are computed by solving an accelerated version of OMP designed to efficiently compute the sparse representations of a batch of signals, called batch orthogonal matching pursuit (OMP) \cite{rubinstein2008efficient}. Unlike convex optimization-based approaches for sparse recovery that use the $\ell_1$-norm \cite{donohoBP}, one can constrain either the total approximation error for each column (Error) or constrain the sparsity (Sparse).

\vspace{-3mm}
\subsection{Variant of Alg.\ \ref{alg:seed}}
We now introduce a variant of SEED that can be used for outlier detection and clustering applications. This variant modifies the way in which we compute the sparse representation of the sampled signals $\X_S$. If we reorder $\X = [ \X_S ~\Xms]$, we can write the sparse matrix $\V = [ \V_S ~\Vms]$, where $\Vms$ contains the sparse representations of the unsampled signals in its columns. Rather than simply setting $\V_S$ to be a diagonal matrix containing the norm of the sampled signals (as in Alg.\ \ref{alg:seed}), we can set $\V_S$ to be the solution of the following objective function:
\begin{equation}
\label{eq:seedvariant}
\min_{\W \in \R^{L \times L}} ~~ \|  \W \|_0 ~~ {\rm s.t.} ~~ \| \X_S - \D \W \|_F \le \epsilon, ~~ {\rm diag}(\W) = 0.
\end{equation}
The idea behind this variant of SEED is to form a sparse representation of each sampled column in terms of other selected columns.

\subsection{Complexity of SEED}
The runtime complexity required to select $L$ columns in Step 1 of SEED (oASIS) is $\mathcal{O}(NL^2)$, requiring storage of a $t \times N$ matrix at step $t$. In Step 2 of SEED, we must compute a sparse approximation of each datapoint, where the runtime complexity of OMP for a $m \times L$ matrix is $T \approx k^3 + kmL$. Thus the complexity of computing $\V$ equals $N( k^3 + kmL)$, which for small $k<L\le m \ll N$, is roughly $\mathcal{O}(NL^2)$. Thus, the total complexity of both steps is given by $\mathcal{O}(NL^2)$. 

To contrast the complexity of SEED with other approaches for clustering, the complexity of computing $k$ nearest neighbors for a collection of $N$ data points is $\mathcal{O}(N^2 \log(N))$. The complexity of SSC-OMP \cite{DyerJMLR13} is dominated by forming a sparse representation of each column of $\X$ with respect to a $m \times N-1$ matrix which has $\mathcal{O}(N^2)$ runtime complexity.


\section{Results for Exact Matrix Recovery}
\label{sec:exactrecovery}
In this Section, we develop sufficient conditions for {\em exact matrix recovery}: this occurs when  the projection of $\X$ onto the subspace spanned by the subset $\X_S$ gives us back exactly $\X$, i.e., $ \X = \proj{S}{\X}$. To prove our main result for exact matrix recovery (Thm.\ \ref{thm:exactrecovery}), we begin by first proving that at each iteration of oASIS, the algorithm selects samples that are linearly independent from those selected at previous iterations (Lem.\ \ref{thm:indep}). We then show how the application of oASIS to the Gram matrix of $\X$ is guaranteed to provide exact matrix recovery.

\subsection{Independent Selection Property of oASIS}
\label{sec:indep}
If we assume that $\X$ is rank $r$, then exact matrix recovery occurs when $\X_S$ contains at least $r$ linearly independent columns from $\X$. In Lemma\ \ref{thm:indep} below, we provide a sufficient condition that describes when oASIS will return a set of $r$ linearly independent columns. 

\begin{LEMM} \label{thm:indep} At each step of oASIS, the $i^{\rm th}$ column of the Gram matrix $\G$ is linearly independent from the previously selected columns provided that $\Delta(i) > 0$.
\end{LEMM}

{\bf Proof.}
We proceed by induction.  Let $\G_S$ denote the set of columns from $\G = \X^T \X$ already selected at the previous iterations and let $\W_k = \X_S^T \X_S$ denote the square matrix consisting of the entries of $\G$ at the selected row and column indices after $k$ columns have been selected. Assume that $\W_k$ is invertible since $S$ consists of linearly independent columns from $\G$. Now, consider selecting the $i^{\rm th}$ column of $\G$ and forming a new $\Wkp$ given by   
\begin{equation*}
 \Wkp = 
 \begin{bmatrix}
\W_k & \bkp \\
\bkp^T & \dkp
\end{bmatrix}, 
\end{equation*}
where $\bkp = \X_S^T \x_i$ is a column vector corresponding to the inner products between the newly selected column $\x_i$ and the previously selected columns (indexed by $S$) and $\dkp = \x_i^T \x_i$ is equal to $\G_{ii}$. This matrix is invertible provided the Schur complement of $\Wkp$ is non-zero.  The Schur complement is $\dkp - \bkp^T \W_k\inv \bkp = \Delta_{k+1}(i).$  Thus, if $ \Delta_{k+1}(i)$ is nonzero, then $\Wkp$ contains linearly independent columns, and thus the $i^{\rm th}$ column of $\G$ from which $\Wkp$ is drawn must also be linearly independent. As long as we initialize oASIS with a set of linearly independent columns, it is guaranteed to select linearly independent columns provided that $\Delta_{k+1}(i) > 0$ for all $i$ corresponding to unselected columns. Our result follows by induction. $\hfill \square$

{\bf Remark.} Lemma \ref{thm:indep} guarantees that oASIS will return a set of $r$ linearly independent columns in $r$ steps as long as the selection criterion $\Delta(i) \ne 0$ holds before exact reconstruction occurs. Unfortunately, in the pathological case in which the algorithm fails with $\Delta(i) = 0$ before $r$ columns have been selected, the algorithm may terminate early.  While it is possible to construct pathological matrices where this occurs, we have not observed this early termination in practice. The following theorem shows that when the entries of the Gram matrix are drawn from a continuous random distribution, the algorithm succeeds with probability $1$.

\begin{THEO}
\label{thm:highprob}
Suppose that the entries of the Gram matrix $\G$ are drawn from a continuous random distribution. Assume that oASIS is initialized by randomly selecting fewer than $r$ columns from $\X$.  Then oASIS succeeds in generating $r$ linearly independent columns with probability $1$.
\end{THEO}

{\bf Proof.} 
We begin by noting that the randomly chosen initialization columns of the Gram matrix $\G$ have full rank with probability $1$.  This is because the matrix $\G$ is a random matrix drawn from a continuous distribution, and the set of singular matrices has positive co-dimension and thus measure 0. The probability of choosing a set of linearly dependent vectors by chance is thus zero.

Suppose now that $k-1$ columns have already been selected, where $k-1<r$.  We wish to show that it is always possible to select column $k.$  The result then follows by induction.

 Observe that the algorithm fails to choose the $k^{\rm th}$ vector only if 
\eqb{whenfail}
\d_i = \b_i^T \W_{k-1}\inv \b_i, \quad \forall i \in \{1,2,\cdots,N\},
\eqe
where  $\d_i$ denotes the diagonal entry in column $i.$  By construction, condition \eqref{whenfail} holds for $i \in \{1,2,\cdots,k-1\}$ (the $k-1$ approximation used by oASIS on iteration $k$ perfectly represents these columns). For columns $k$ through $r,$ the quantity $\b_i^T \W_{k-1}\inv \b_i,$ is known (i.e., it is not a random variable because it can be computed using only values of the sampled columns $1$ through $k-1$).  However, for such columns, $\d_i$ is  continuous random variable, and thus the probability of \eqref{whenfail} holding for $i\ge k$ is 0. $\hfill \square$

\subsection{Exact Matrix Recovery}
Using Lemma \ref{thm:indep}, we now prove that oASIS returns a sample subset that yields exact matrix recovery. To do this, we use the fact that the Gram matrix $\G = \X^T \X$ from which oASIS selects columns spans the same space as $\X$. Thus, when we select $r$ linearly independent columns from $\G$, this also guarantees that the same set of columns from $\X$ are linearly independent. This idea is made precise in the following.
\begin{LEMM}
\label{lemma:exactrecovery}
Let $\X$ be a rank $r$ matrix and $\G_S$ be the set of columns selected via oASIS from the corresponding Gram matrix. Exact matrix recovery of $\X$ occurs when ${\rm rank}(\G_S) = r$.
\end{LEMM}

{\bf Proof.}
Recall that the Gram matrix $\G = \X^T \X$ is by definition of the same rank as $\X$. The columns of $\G$ indexed by $S$ equals $\G_S=\X^T\X_S$ and the ${\rm rank}( \G_S)= {\rm rank}(\X^T\X_S) = {\rm rank}(\X_S).$ Thus, when ${\rm rank}(\G_S) = r$, this implies that ${\rm rank}(\X_S) = r$, which by assumption equals the rank of the full dataset $\X$. Therefore, when ${\rm rank}(\G_S) = r$, exact recovery of $\X$ is guaranteed. $\hfill \square$
\newtheorem{ASM}{Assumption}

To state our main result for matrix recovery with oASIS, we must make the following assumption. 
\begin{ASM}
\label{asm:asm1}
For a rank $r$ matrix, oASIS returns a set of $r$ columns before terminating with $\max_i = \Delta(i) = 0$. 
\end{ASM}
We are now equipped to state our main result.
\begin{THEO}[Exact Recovery Condition]
\label{thm:exactrecovery}
Let $\X$ be a rank $r$ matrix. Exact matrix recovery occurs as long as Assumption \ref{asm:asm1} is satisfied.
\end{THEO}

{\bf Proof.}
To prove Thm. \ref{thm:exactrecovery}, we must simply combine Lemma \ref{thm:indep} with Lemma \ref{lemma:exactrecovery}. To be precise, Lemma \ref{thm:indep} states that oASIS will return a set of $r$ linearly independent columns from $\G$ provided that $\Delta(i) > 0$. Thus, as long as the algorithm does not terminate with $\Delta(i) = 0$, this implies that we return a set of $r$ linearly independent columns from $\G$ indexed by the index set $S$. Now, using Lemma \ref{lemma:exactrecovery}, we have that when oASIS returns a subset of columns $\G_S$ such that the ${\rm rank} (\G_S) = r$, then exact recovery is guaranteed for $\X$ based upon the corresponding subset $\X_S$. $\hfill \square$

\subsection{Exact Recovery from Unions of Independent Subspaces}
\label{sec:unionsubs}
Guarantees for EFS for SSC \cite{soltanolkotabi2012geometric} and SSC-OMP \cite{DyerJMLR13} rely on the assumption that the dataset $\X$, provides a {\em complete reference set} for each low-dimensional subspace present in the data (see Sec.\ \ref{sec:ssc}). To make this precise, assume that the points in $\X$ are drawn from a union of $p$ subspaces, where $\mathcal{U} = \cup_{i=1}^p \sub_i$ and the ${\rm dim}(\sub_i) = k_i$. We say that $\X$ provides a complete reference set for $\mathcal{U}$ if $\X$ contains at least $k_i$ points from $\sub_i$ for all $i = \{1, \dots, p\}$.

It follows from the definition of exact matrix recovery that, if $\mathcal{U}$ is a union of independent subspaces and ${\rm dim}(\X) = \sum_{i} k_i$, then whenever $\X_S$ yields exact matrix recovery we are guaranteed that $\X_S$ also provides a complete reference set for $\mathcal{U}$. This is the main condition required to prove EFS in \cite{DyerJMLR13,soltanolkotabi2012geometric}; thus, when SEED returns a subset of the data with at least $k$ columns for each of its $k$-dimensional subspaces, we can compute the corresponding covering radius of each subspace (see \cite{DyerJMLR13} for further details). Thus, as long as exact recovery occurs, we can apply the theory in \cite{DyerJMLR13,soltanolkotabi2012geometric} to produce guarantees that EFS occurs for the decomposition obtained via SEED. This result follows from combining Thm.\ 1 in  \cite{DyerJMLR13} with our condition for exact recovery in Thm.\ \ref{thm:exactrecovery}.

\begin{figure*}[t!]
        \centering
                \includegraphics[width=\textwidth]{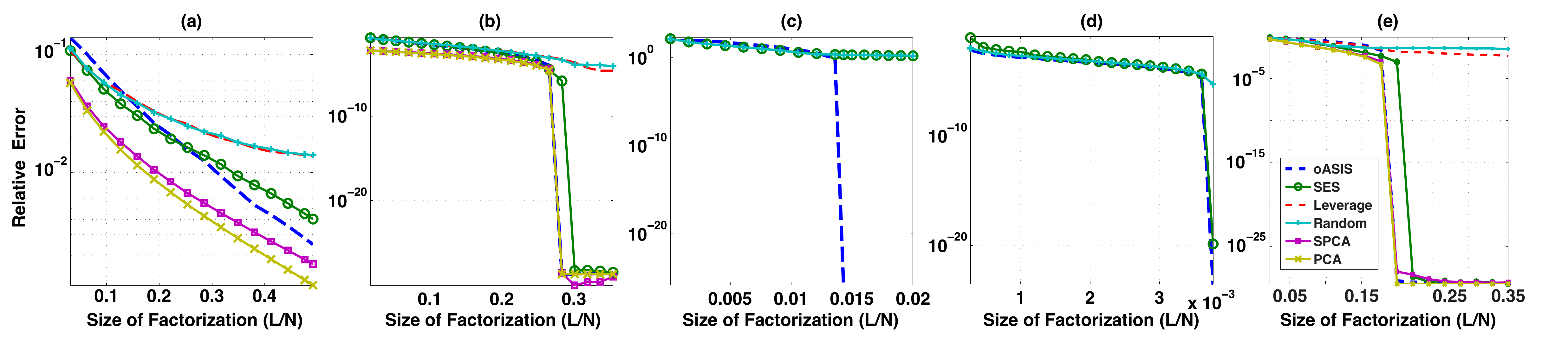}
                \caption{{\bf Approximation error versus size of factorization.} {The relative approximation error is displayed as a function of the factorization size ($L/N$) for SEED, error-based sampling (SES),  random sampling for: (a) Faces , (b) Neuro, (c) MNIST, (d) HS, and (e) UoS. For the smaller datasets in (a,b,e), we also display the error for leverage sampling, PCA, and SPCA.}
                 \vspace{-5mm} }
                \label{fig:mainfig}
 \end{figure*}

\section{Numerical Experiments}
In this Section, we first introduce the datasets used in our evaluations and then evaluate the performance of SEED for matrix approximation, clustering, denoising, and outlier detection.

\label{sec:results}

\subsection{Datasets and Evaluation Setup}
\label{sec:datasets}
We now describe the datasets used in our evaluations.
\begin{itemize}
\item The Face dataset consists of images (each with $4032$ pixels) of ten subjects faces under various illumination conditions, resulting in a dataset of size $4032 \times 631$ \cite{yaleb}. 
\item The hyperspectral (HS) dataset consists of $204$ images from the Salinas scene, where each image contains spatial information about the scene at a different spectral band. Each image is $512 \times 217$ pixels and after removing pixels without labels, the total dataset is of size $204 \times 54129$. This dataset consists of $16$ types of vegetation (classes) and the spectral signatures associated with each class are very low-dimensional (approximately $3$--$10$ dimensions). 
\item The MNIST dataset contains $50,000$ images of $10$ handwritten digits ($28 \times 28$ pixels), which results in a dataset of size $784 \times 50k$ \cite{lecun95b}. 
\item The Neuro dataset consists of the firing rates of $187$ neurons in motor area (M1) collected at $N = 875$ time points, when a monkey is performing a center-out reach task moving from a center position to one of $8$ targets \cite{stevenson2011statistical}. The resulting dataset produces a data matrix of size $187 \times 875$. 
\item The union-of-subspaces UoS dataset is a synthetic dataset consisting of signals drawn from a union of two subspaces of dimension $k=20$ with a $3$-dimensional overlap (three coordinates are fully shared by both subspaces). We then add a collection of outliers created by generating random Gaussian vectors. This results in a dataset of size $N = 450$ with $N_1 = 300$ points in the first subspace, $N_2 = 100$ points in the second subspace, and $N_o = 50$ outlier points.
\end{itemize}

For our evaluations on the HS and MNIST datasets, we used an OpenMPI implementation of SEED written in C++. This parallelizes both oASIS for faster column selection and batch OMP for faster sparse representation computation. The experiments utilized 72 total processor cores with 4 GB of RAM per core. For the smaller datasets, we ran all of our evaluations in MATLAB with a single desktop processor. 
 
\subsection{Matrix Approximation}
\label{sec:mappx}
A well-studied application of CSS is the approximation of low rank matrices \cite{mahoney2009cur}. As SEED utilizes a fast sequential algorithm for column selection (oASIS), our approach provides an effective strategy for matrix approximation. As we will show in Sec.\ \ref{sec:indep}, oASIS selects linearly independent columns at each step and thus obtains exact recovery (to machine precision) of rank $r$ matrices using $r$ samples (this is observed in practice in Fig.\ \ref{fig:mainfig}). In contrast, random and leverage-based sampling exhibit a significantly slower decay in their approximation error. 

\subsubsection{Results for Matrix Approximation}
To evaluate the performance of SEED for matrix approximation, we compute the approximation error as a function of decomposition size. We compare the error when sampling the dataset with: (i) oASIS, (ii) sequential error selection (SES) \cite{deshpande2006matrix},  (iii) uniform random sampling, and when possible (iv) leverage score sampling \cite{mahoney2009cur}. In addition, we also compute the error for PCA and SPCA using the generalized power method in \cite{journee2010generalized}. The relative approximation error is $${\rm err}(\X,\D) = \frac{\| \X - \D \D^{+} \X \|_F^2}{\| \X \|_F^2}.$$

Figure \ref{fig:mainfig} displays the approximation error for all five datasets as a function the relative factorization size $L/N.$  oASIS achieves exact matrix recovery for (b-e) when the number of points sampled equals the rank. This is in agreement with our result for matrix recovery in Thm.\ \ref{thm:exactrecovery}. This result suggests that our guarantee for exact matrix recovery is not overly restrictive and that SEED produces linearly independent sample sets in a wide range of real and synthetic datasets.

Interestingly, we observe similar decay in the approximation error for both the Neuro and synthetic UoS datasets: (i) the error achieved for SEED, PCA, and SPCA are roughly equivalent (all three achieve exact recovery with the number of factors/samples equal to the rank), (ii) SES trails behind SEED but also quickly achieves exact recovery, and (iii) random and leverage sampling flatline and do not achieve exact recovery. This suggests that the Neuro dataset is likely to contain both low rank structures as well as outliers that make random sampling significantly less effective for matrix approximation than SEED.

\subsection{Sparse Representation-Based Learning}
\label{sec:clustering}
Sparse representation-based approaches to classification \cite{wright2010sparse} and clustering \cite{elhamifar2009sparse}, leverage the fact that signals from the same class (or subspace) will use one another in their sparse representations. Using this fact, the sparsity patterns of a self-expression decomposition such as SSC can be used to cluster the data using either a spectral clustering or consensus method \cite{gowreesunker2010learning} (see Sec.\ \ref{sec:ssc} for more details). In fact, recent studies have shown that, when the dataset lives on a union of subspaces, the sparse representation of a point from a subspace will only consist of points from the same subspace \cite{elhamifar2009sparse,vidaljournal,soltanolkotabi2012geometric,wang2013noisy,DyerJMLR13}.

Just as in SSC, we can use the decomposition provided by SEED to cluster the data using the sparsity patterns in $\V$. However, because $\V$ is rectangular, standard spectral clustering approaches for square affinity matrices cannot be used. Rather, we can think of $\V$ as representing the edges of a bi-partite graph and thus co-clustering methods can be used in place of standard graph clustering methods. The spectral co-clustering algorithm introduced in \cite{dhillon2001co} provides an elegant relaxation of the problem of finding a minimum cut through a bi-partite graph. An interesting consequence of using a spectral co-clustering approach is that, when we eventually solve an eigenvalue problem to find the minimum cut, we compute the second {\em largest} singular vector of a $L \times N$ matrix rather than the second {\em smallest} eigenvector of a $N \times N$ matrix. This enables us to exploit simple iterative methods for leading singular vectors rather than computing the entire SVD for a $N \times N$ matrix.

\subsubsection{Results for Sparse Representation-Based Learning}

In Figure \ref{fig:coclust}, we show a visualization of the embedding of a union of five overlapping subspaces; we show the first three coordinates of the embedding and plot the projection of the unsampled signals as red dots and the projection of sampled signals as blue stars. This result provides evidence that: (i) co-clustering provides a feasible and efficient strategy for clustering the data with SEED and (ii) our proposed sampling strategy is capable of separating the data with far fewer samples than random sampling. 

To quantify the performance of SEED for sparse representation-based clustering, we compute the cost of a normalized cut as we vary the size of the factorization $L$. The cost of a normalized cut is a measure of how easy it is to cluster a bi-partite graph into its correct classes. This cost is defined as follows \cite{dhillon2001co}: Let $R_k$ index the rows of $\V$ corresponding to points in the $k^{\rm th}$ class, $C_k$ index the set of columns corresponding to the points in the $k^{\rm th}$ class, and $\Omega$ index all $N$ points in the dataset. The cost of the normalized cut for the $k^{\rm th}$ class is
$${\rm ncut}(\V) =  \frac{\sum_{i \in {R_k}, j \in {C_k}} | \V_{ij} | }{  \sum_{i \in R_k, j \in \Omega} | \V_{ij} | } + \frac{\sum_{i \in {R_k}, j \in {C_k}} | \V_{ij} | }{  \sum_{i \in \Omega, j \in {C_k}} | \V_{ij} | }.$$
In our subsequent experiments, we compare the average cost of a normalized cut for column sampling-based approaches with SSC-OMP and the NN graph. We note that both SSC-OMP and NN exhibit $\mathcal{O}(N^2)$ complexity, which make them impractical for large datasets.

For the Neuro dataset, we achieve lower cut ratios than SSC ($0.1755$) and NN ($0.1613$) for all of the column sampling-based approaches after sampling $20\%$ of the samples. With $30\%$ of the samples, the column sampling-based methods achieve cut ratios of $0.1085$ (SEED), $0.1342$ (SES), $0.1512$ (Lev), $0.1412$ (Rand). For these experiments, we set $k_{\rm max} = 5$ and $\epsilon = 0.7$. For higher values of $k_{\rm max}$ and lower values of $\epsilon$, we observe that SSC-OMP provides the best cut ratios. Our results suggest that, while leverage and random sampling provide poor schemes for matrix approximation (as evidenced by Fig.\ \ref{fig:mainfig}), all of the column sampling approaches provide comparable performance in terms of their cut ratios.

\begin{figure}[t!]
\centering
\label{fig:coclust}
\centerline{\includegraphics[width=0.75\columnwidth]{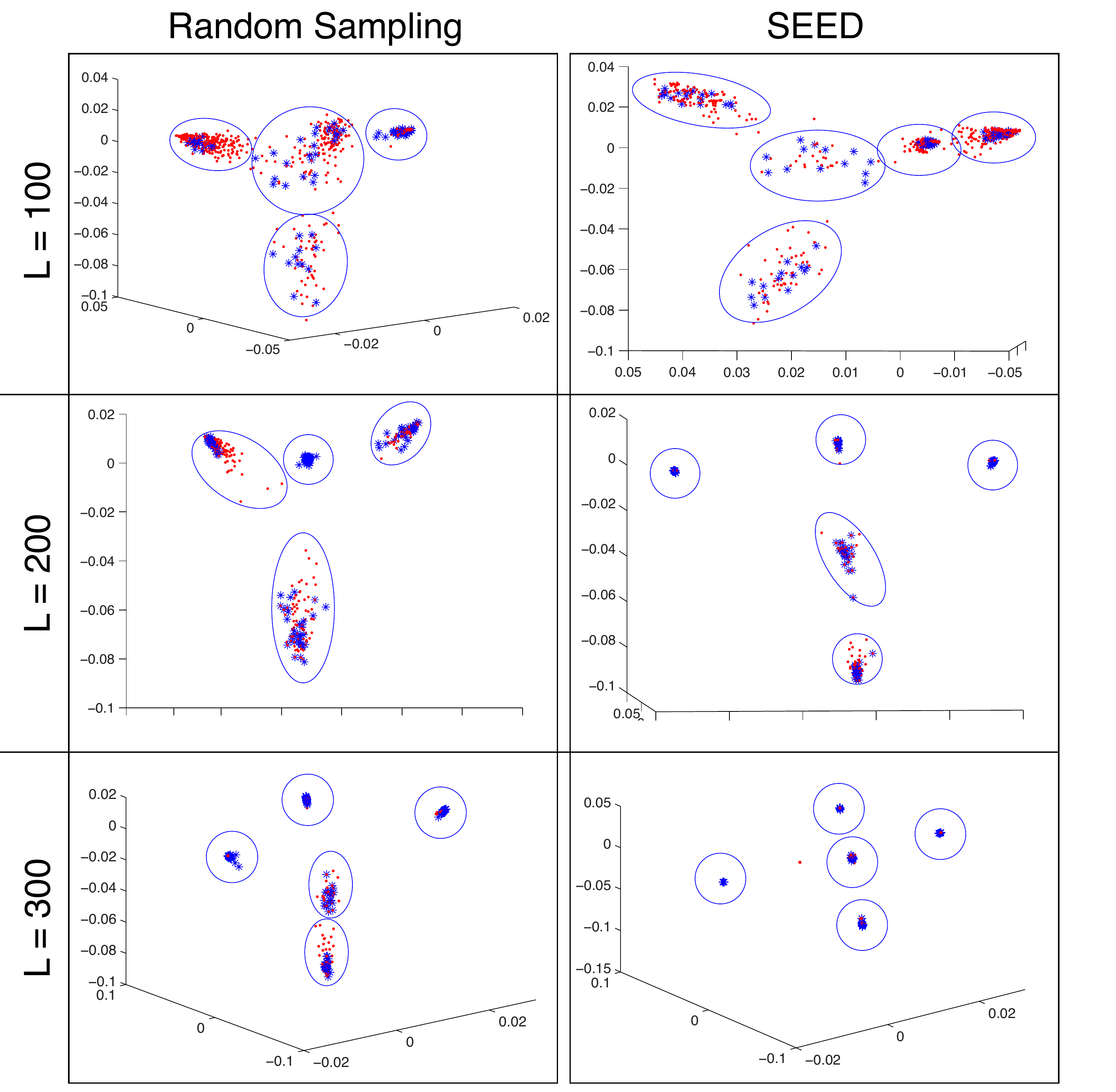}}
\vspace{-0.2cm}
\caption{{\bf Visualization of co-clustering with SEED.} Visualization of the embedding of $\V$ for a union of five overlapping $20$-dimensional subspaces in $\R^{200}$. Pairs of subspaces have at most a $10$-dimensional intersection and the rank of the dataset is $r = 150$. On the left, we show the embedding for $L = \{100, 200, 300\}$ samples selected via random sampling and on the right, we show the embedding for SEED. To aid in visualization, we draw ellipses around samples from each subspace (each cluster) and display the sampled points as blue ($\ast$) and unsampled points as red ($\circ$). }\label{fig:coclust}
\vspace{-5mm}
\end{figure}

\begin{figure}[t!]
        \centering
                \includegraphics[width=\columnwidth]{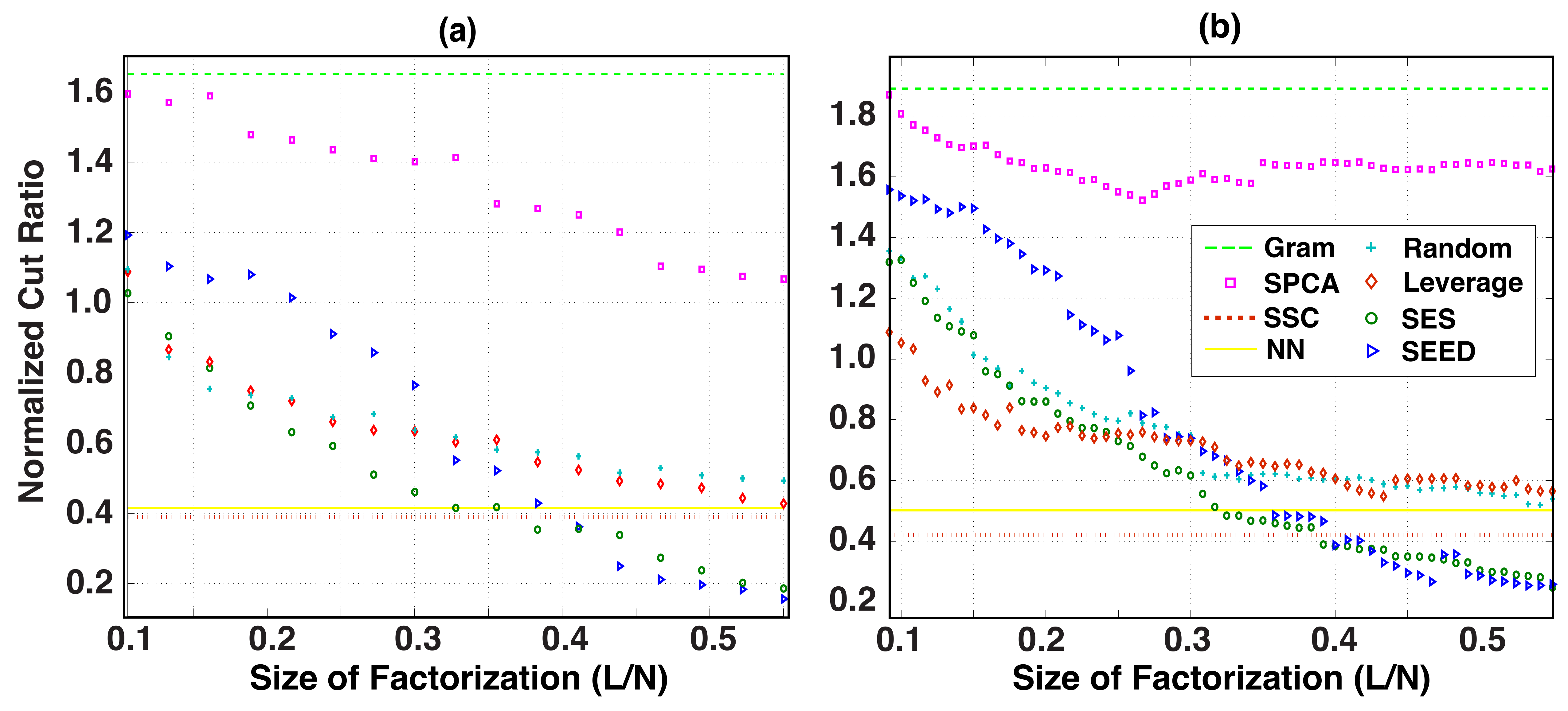}
                \caption[{\bf Normalized cut ratios for face image database.} ]{ \label{fig:ncut_faces} {\bf Normalized cut ratios for face image database.} Normalized cut ratios vs. size of factorization for collections of (a) six subject's faces under sixty different illumination conditions ($M = 4032, N = 360$) and (b) twenty subject's faces under sixty different illumination conditions ($M = 4032, N = 1200$). In both cases, the data is full rank, i.e., $r = 360$ and $r = 1200$ for (a) and (b), respectively.}
                \label{fig:facencuts}
                \vspace{-3mm}
 \end{figure}

In Fig.\ \ref{fig:facencuts}, we display the normalized cut ratios for the Faces dataset for six (a) and twenty (b) different subjects. We observe similar decay in the cut ratios for both datasets, where SEED and SES achieve normalized cuts less than NN and SSC-OMP when we sample $35\%$ of the dataset. The gap between SSC-OMP and SEED grows as we increase the number of samples to $50\%$. The performance of leverage and random sampling appears to flatline just above the cut ratio for SSC and NN methods. 

In many of the datasets that we tested, we observe that subsampling-based approaches can produce smaller cut ratios than SSC-OMP and NN methods. This is likely due to the fact that, as the size of the ensemble grows relative to the dimension of the underlying cluster, the graphs generated by SSC and NN become weakly connected and thus produce smaller cut ratios \cite{DyerICASSP2013,nasihatkon2011graph}. In contrast, the sparse representations produced by SEED are built upon self-expressive bases containing incoherent columns, and thus we observe that SEED produces sparse graphs that are more well connected (within cluster) and thus produce smaller cut ratios (Sec.\ \ref{sec:results}). 

\subsection{Denoising}
\label{sec:denoising}
SEED computes a sparse approximation of each column of $\X$ in terms of a small number of columns from the same dataset. By constraining the sparsity level of each column of $\V$ in OMP (solving Sparse), we obtain an approximation to the $i^{\rm th}$ column as $\widehat{\x}_i = \X_S \v_i$. This approximation scheme provides a denoised version of the original dataset that is similar in spirit to NN-based denoising. However, rather than finding the $k$ nearest neighbors and applying a simple averaging procedure (NN-denoising), SEED finds both an optimal set of ``neighbors'' and weights to use at each datapoint. As a motivating example (Fig.\ \ref{fig:hsiresult}), we show the performance of SEED for clustering hyperspectral image data after denoising the data with (d) SEED, (c) a random subset of samples, and after (b) applying $k$-means to the original data. Here, we observe a significant improvement in clustering after denoising the data with only $L = 30$ samples: the $k$-means clustering error before and after denoising the data with SEED is $31\%$ and $0.68\%$ respectively.

\subsubsection{Results for Denoising}
\begin{figure}[t!]
\centering
\centerline{\includegraphics[width=1.04\columnwidth]{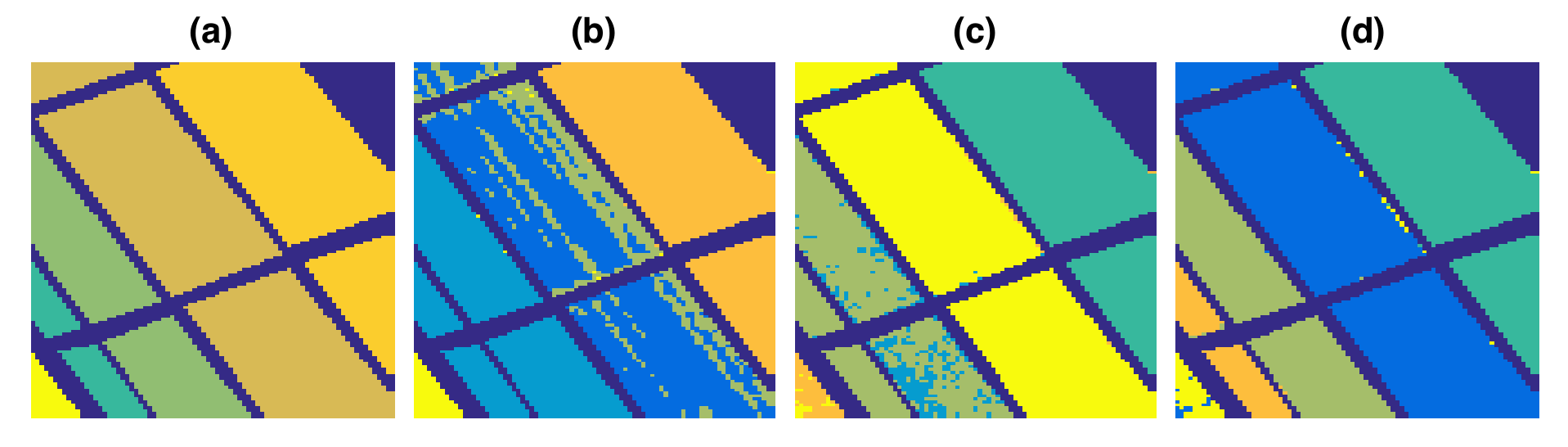}}
\caption{{\bf Clustering hyperspectral images (HSI).} We display the results of clustering a section of the HSI data: (a) ground truth (b) and $k$-means applied to the original data, (c) denoised data from a random selection of $L = 30$ columns, and (d) denoised data obtained via SEED for $L = 30$ columns. The clustering error is $31\%$ (b), $15\%$ (c), and $0.68\%$ (d). }\label{fig:hsiresult}
\end{figure}

Due to the noisy nature of hyperspectral data, we find that SEED is a highly effective strategy for denoising such datasets. To do this, we apply SEED to compute the approximation $\widehat{\X} = \D \V$, where $k_{\rm max} = 5$ and the error tolerance for OMP to $\epsilon = 0.2$. After denoising the data, we then apply a simple $k$-means clustering algorithm to the columns in $\widehat{\X}$. In Fig.\ \ref{fig:hsiresult}, we display the results from clustering a subset of the image with only $L = 30$ points selected from the entire dataset ($N = 54129$). With only a small sample of points, we observe nearly perfect clustering of the image with SEED. We compare the performance of SEED with a random sampling-based approach (identical to the setup for SEED except the sample sets are selected randomly) and clustering the original data without denoising. The clustering error is $31\%$ ($k$-means applied to original data), $15\%$ (denoising with random samples from dataset), and $0.68\%$ (SEED). While we do not show the clustering results obtained by denoising the data with PCA, the clustering error for PCA-based denoising based upon $L = 30$ principal components is $62\%$ (significantly higher than clustering the raw data).

\subsection{Outlier Detection}
\label{sec:outliers}
When the dataset lies on a single or multiple low-dimensional subspaces, we can use SEED to discover outliers. The idea behind using SEED for outlier detection is that we try to sparely represent a data point that lies in a low-dimensional subspace, a small number of data points are required (i.e., the representation is sparse). In contrast, when we try to sparsely represent an outlier, a large number of data points are needed (i.e., the representation is dense). For instance, when a collection of signals lie on a union of $k$-dimensional subspaces, the sparsity of each signal in a $k$-dimensional subspace is bounded by $k$. We can exploit this {\em rank revealing} property of SEED to determine whether a signal lies on one of the low-dimensional subspaces in the ensemble or whether it is an outlier.

To find outliers in the dataset, we form a self-expressive basis and then form the decomposition $\widehat{\X} = \D \V$, where $\V = [\V_S ~ \Vms]$. Rather than setting $\V_S$ to a diagonal matrix as in Alg.\ \ref{alg:seed}, the sparse coefficients $\V_S$ are obtained by solving the SSC objective in (\ref{eq:seedvariant}). To compute both $\V_S$ and $\V_{-S}$, we utilize batch OMP to solve (Error) by providing an error tolerance $\epsilon$ to the algorithm. Constraining the error (rather than the sparsity) is important, because our goal is to use the sparsity level of each column to determine whether it is an outlier. Once we compute a sparse factorization, we compute the number of nonzeros in each column of $\V$ (sparsity level)  and segment the columns of $\X$ based upon a user set threshold. When a column of $\V$ is dense we declare it an outlier and when $\V$ is sufficiently sparse, we declare it an inlier. In some cases, setting a threshold to segment the data can be straightforward (when the sparsity levels admit a bi- or multi-modal distribution). However, in cases where setting the threshold is difficult, one can use $k$-means to learn a threshold to split the data. 

\subsubsection{Results for Outlier Detection}
In Fig.\ \ref{fig:outlierdemo}, we demonstrate the rank revealing property of SEED when applied to the UoS dataset that has been corrupted with outliers. Along the bottom, we show the sparse coefficient matrices obtained for (a) SEED ($L=160$), (b) Random sampling ($L=160$), and (c) SPCA ($L=60$). Above these coefficient matrices, we show the number of nonzeros (sparsity) of each column. In the case of SPCA, we set $L = 60$ because, as we increase $L$, we observe an even smaller gap in the sparsity level between signals living in low-dimensional subspaces and the outliers. For all of the methods, we compute the sparse coefficients using OMP, where we set $\epsilon = 0.3$ and do not constrain the maximum sparsity level. In this case, we can clearly separate inliers from outliers: the sparsity level of inliers and outliers is around $k=2$ and $k =10$, respectively. 

Our results suggest that SEED can provide a strategy for outlier detection by simply thresholding columns based upon their sparsity level. In general, determining an appropriate threshold to segment low rank structures from outliers can be challenging. However, in practice, we observe that the distribution of column sparsity is multi-modal; thus, instead of setting the threshold explicitly, a $k$-means algorithm can be employed to find a good threshold to segment outliers.

 \begin{figure}[t!]
\centering
\label{fig:outlierdemo}
\centerline{\includegraphics[width=\columnwidth]{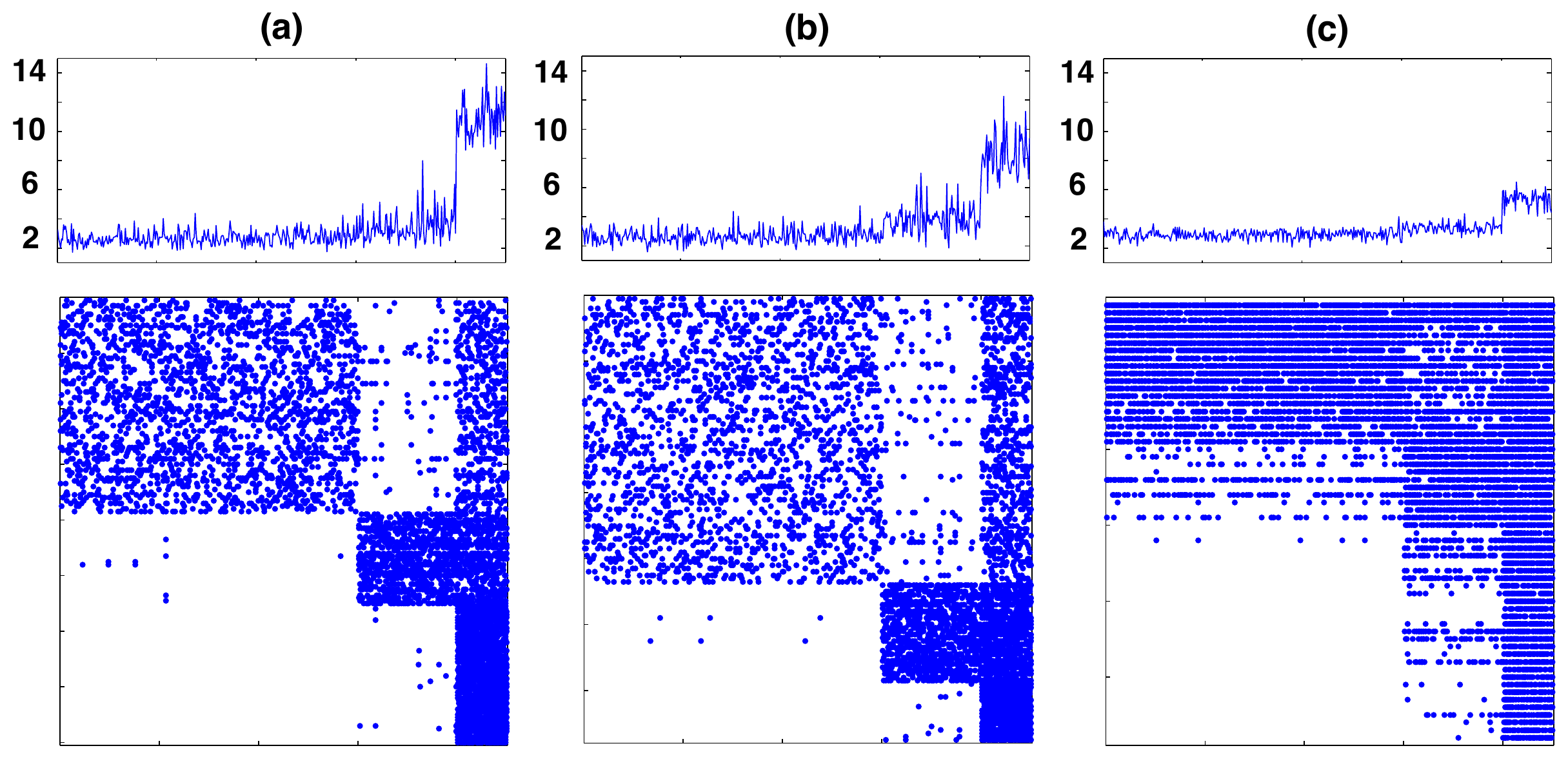}}
\vspace{-0.2cm}
\caption[{\bf Outlier detection with SEED.}] {{\bf Demonstration of rank revealing property of SEED.} The sparsity level (top row) and sparse coefficient matrices (bottom row) for SEED with (a) SEED ($L=160$), (b) Random sampling ($L=160$), and (c) SPCA ($L=60$). \label{fig:outlierdemo}}
\end{figure}

\section{Conclusions}
\label{sec:conclusion}
This paper introduced SEED, a scalable method for sparse matrix factorization that couples a new and provable method for column selection (oASIS) with greedy sparse recovery (OMP).  We have demonstrated how SEED can be applied to either assist or solve numerous signal processing and machine learning problems, ranging from matrix approximation and denoising, to clustering and outlier detection. In addition, we have developed a sufficient condition for SEED to achieve exact matrix recovery when we sample the same number of columns as the rank of the dataset (Thm.\ \ref{thm:exactrecovery}). In numerical experiments, we have shown that this result holds for a number of real-world datasets, i.e., we obtain exact recovery after sampling a number of columns equal to the matrix rank. This is in stark contrast to random sampling, where exact recovery cannot be guaranteed. 

Column sampling has been explored extensively in the machine learning literature for the task of approximating low rank matrices \cite{Williams01usingthe,journals/jmlr/DrineasM05,deshpande2006matrix}. In this paper, we applied column selection to a new class of problems, namely sparse representation-based clustering/classification and subspace clustering. Thus, an important contribution of our work is showing how self-expressive approaches used in signal processing and computer vision can benefit from column selection approaches.

We have demonstrated that SEED provides self-expressive bases amenable to solving sparse representation-based learning and subspace clustering. In the case where the dataset lies on a union of independent subspaces, we have shown that our condition for exact recovery (Thm.\ \ref{thm:exactrecovery}) also implies that we select at least $k$ linearly independent columns from each $k$-dimensional subspace in the dataset. However, providing a bound on the covering radius (how well each subspace is sampled) with column sampling methods is an open problem that must be solved to prove stronger results for feature selection similar to those in \cite{vidaljournal, soltanolkotabi2012geometric, DyerJMLR13} for SSC. Extending our analysis to the case of approximately low rank matrices, unions of overlapping subspaces, and noisy settings are all interesting directions for future work.

\appendices
\section{Accelerated Sequential Incoherent Selection (oASIS)}
We now provide a detailed description of our implementation of oASIS for column sampling. Pseudocode is provided in Alg.\ \ref{alg:oasis}.

 \subsection{Accelerated Sampling}
A na\"ive implementation of the column sampling approach described in Sec.\ \ref{sec:sis} is inefficient, because each step requires a matrix inversion to form $\Wkp^\dagger$ in addition to calculating the errors $\Delta_i.$  Fortunately, this can be done efficiently by updating the results from the previous step using block matrix inversion formulas and rank-1 updates. We now provide a derivation of the algorithm and pseudocode in Alg.\ \ref{alg:oasis}.

\begin{algorithm}[t!]
   \caption{{\bf : Accelerated Sequential Incoherent Selection (oASIS)}}
   \label{alg:oasis}
\begin{algorithmic}
	\STATE{\bfseries Inputs:} The data matrix $\X \in \R^{m \times N}$, a $N \times 1$ vector $\d = {\rm diag}(\G)$, the maximum number of columns to sample $L$, the number of columns to sample initially $k<L$, and a non-negative stopping criterion $\delta.$ 
	\STATE{\bfseries Initialize: } Choose a vector $S \in [1,N]^K$ of $k$ random starting indices. Set $\C = \X^T\X_S$ , $\W\inv = {(\X_S^T\X_S)}\inv$, $\d = {\rm diag}(\X^T \X_S)$ and ${\bf R} = \W\inv\C^T$.
	\WHILE{$k<L$} 
	\STATE $\Delta \leftarrow \d - \colsum(\C \circ {\bf R})$
	\STATE $i \leftarrow \argmax_{j \not\in S}  |\Delta|$
	\IF{$|\Delta_i|<\delta$}
	\STATE {\bfseries return}
	\ENDIF 
	\STATE $\b \leftarrow \X_S^T\x_i$ , $\s \leftarrow 1/\Delta_i$ ,  ${\bf q} \leftarrow {\bf R}_i$, $\c \leftarrow \X^T\x_i$
	\STATE $\C \leftarrow [\C,  \c]$
	\STATE $\W\inv \leftarrow 
		\begin{bmatrix}
		\W\inv+\s{\bf q}{\bf q}^T, 	& 	-\s{\bf q}  \\
		-\s{\bf q}^T,		& 	\s
		\end{bmatrix}$
	\STATE ${\bf R} \leftarrow 
		\begin{bmatrix}
		{\bf R}+ \s {\bf q}  ({\bf q}^T \C^T-\c^T) \\
		\s(-{\bf q}^T \C^T+\c^T)
		\end{bmatrix}$
 \STATE $k\leftarrow k+1$
 \STATE $S \leftarrow S \cup \{i\}$
	\ENDWHILE
	\end{algorithmic}
\end{algorithm}

We first consider the calculation of $\Wkp^\dagger$ after column $k+1$ is added to the approximation.  Let $\b_k$ denote the first $k$ rows of column of the $k+1^{\rm th}$ column of $\G$ and $\d_k$ denote its diagonal.  Using a block inversion formula, we obtain
 \begin{align}
 \label{eq:update1}
 \Wkp\inv &= 
 \begin{bmatrix}
\W_k & \bkp \bkp^T &\dkp
\end{bmatrix}\inv \\
 &=  
\begin{bmatrix}
 \W\inv_k+\skp \qkp \qkp^T, 	& 	- \skp \qkp  \\
 -\skp \qkp^T,		& 	\skp
 \end{bmatrix},
\end{align}
where $\skp = (\dkp - \bkp^T \W\inv\kp \bkp)\inv = \Delta\kp\inv $ is the (scalar valued) Schur complement and $\qkp=\W\inv_k \bkp$ is a column vector.  This update formula allows $\Wkp\inv$ to be formed by updating $\W_k\inv,$  and only requires inexpensive vector-vector multiplication.  Note that $\Wkp$ is invertible as long as $\Delta\kp$ (the Schur complement) is non-zero, which is guaranteed by our sampling rule:  the algorithm terminates if $\Delta\kp = 0$ in which case our approximation is exact.

We now consider the calculation of $\Delta_i= \d_i - \b_i^T\W_k^\dagger \b_i $ for all $i.$    Note that on step $k$ of the method, we have $\C_k^T=[ \b_1, \b_2,\cdots,\b_N].$ We can evaluate all values of $\b_i^T\W_k^\dagger \b_i $ simultaneously by computing the entry-wise product of $\C_k$ with the matrix ${\bf R}_k = \W\inv_k \C_k^T$ and then summing the resulting columns.  
   If we have already formed $\C_k$ and ${\bf R}_k$ on iteration $k,$ then the matrix ${\bf R}\kp = \W\inv\kp \Ckp^T$ needed on the next iteration is obtained by applying Eqn.\ \ref{eq:update1} to $\Ckp^T$ to obtain 
\begin{align*}
\label{eq:update2}
 {\bf R}\kp &= \Wkp\inv 
 \Ckp^T
 =
\Wkp\inv 
  \begin{bmatrix}
\C^T_k \\
\ckp^T
 \end{bmatrix}\\
&=
 \begin{bmatrix}
{\bf R}_k+ \skp \qkp  (\qkp^T \C^T_k- \ckp^T) \\
\skp(- \qkp^T \C^T+ \ckp^T)
 \end{bmatrix}.
\end{align*}
The update formula above forms ${\bf R}\kp$ by updating the matrix ${\bf R}_k$ from the previous iteration.  The update requires only matrix-vector and vector-vector products.   The application of this fast update rule to perform incoherent sampling yields Alg.\ \ref{alg:oasis}. We use this accelerated version of oASIS in all of our numerical expeditions.

\section*{Acknowledgment}
The authors would like to thank Azalia Mirhoseini, Ebrahim Songhori, and Farinaz Koushanfar for helpful discussions and their assistance in developing MPI code for running SEED on large datasets. Thanks also to Matt Perich, Lee Miller, and Mohammad Azar for collecting and sharing the neural data used in our evaluations. ELD, TAG, RJP, and RGB were funded by NSF CCF-0926127, NSF CCF-1117939, ONR N00014-12-1-0579, ONR N00014-11-1-0714, and ARO MURI W911NF-09-1-0383. ELD and KPK were funded by R01MH103910. ELD was also funded by NSF GRFP 0940902 and a Texas Instruments Distinguished Graduate Fellowship. 

\ifCLASSOPTIONcaptionsoff
  \newpage
\fi

\bibliography{seed2015}
\bibliographystyle{ieeetr}

\end{document}